\documentclass[pdftex,twocolumn,epjc3]{svjour3}          % twocolumn

\RequirePackage[T1]{fontenc}

\smartqed  % flush right qed marks, e.g. at end of proof

\RequirePackage{graphicx}
\RequirePackage{mathptmx}      % use Times fonts if available on your TeX system
\RequirePackage{flushend}
\RequirePackage{xspace}
\RequirePackage{amsmath}
\RequirePackage[numbers,sort&compress]{natbib}
\RequirePackage[colorlinks,citecolor=blue,urlcolor=blue,linkcolor=blue]{hyperref}

\journalname{Eur. Phys. J. C}

\newcommand{\pythia}{\textsc{Pythia~8.3}\xspace}
\newcommand{\epos}{\textsc{Epos4}\xspace}
\newcommand{\epp}{$e^+e^-$\xspace}
\newcommand{\pp}{$\mathrm{pp}$\xspace}

\begin{document}

\title{A test of strangeness quantum number conservation \\ in proton-proton collisions}

\author{Christian Bierlich\thanksref{e1,addr1}
        \and
        Stefano Cannito\thanksref{e2,addr2} 
        \and 
        Valentina Zaccolo\thanksref{e3,addr2} 
}

\thankstext{e1}{e-mail: christian.bierlich@fysik.lu.se}
\thankstext{e2}{e-mail: stefano.cannito@studenti.units.it}
\thankstext{e3}{e-mail: valentina.zaccolo@units.it}

\institute{Department of Physics, Lund University, Box 118, Lund, SE-221 00, Sweden\label{addr1}
\and
Department of Physics, University of Trieste, Via Alfonso Valerio 2, Trieste, 34127, Italy\label{addr2}
}

\date{Received: date / Accepted: date}

\maketitle

\begin{abstract}
    The study delves into the production of \mbox{(multi-)} strange hadrons in proton-proton collisions at LHC. Novel observables are proposed to distinguish between \epos,  
    based on core-corona separation between a thermalised QGP phase and a vacuum phase with global strangeness conservation, and \pythia, based on microscopic interactions between Lund strings that conserve strangeness locally. Correlations between a $\phi$ meson and (multi-)strange hadrons are shown to be an excellent discriminator between the two types of models.
\end{abstract}

\section{Introduction}
Hadronisation is the complex process of the formation of hadrons from partons. Due to its non-perturbative nature, no exact theory exists and one has to rely on phenomenological assumptions to describe it.  Historically, hadronisation has been treated differently at different density regimes and a good proxy for density is the multiplicity of produced particles. At low multiplicity the produced particles are few, the system is dilute (vacuum like), while at high multiplicity the system produced is dense. The most dense environment can be reached in collisions of heavy nuclei (AA). 

As the system gets denser, enhanced production of \mbox{(multi-)} strange hadrons with respect to the \epp baseline is observed and this feature is assumed to be a signal of Quark--Gluon Plasma (QGP) production \cite{Cabibbo:1975ig, Shuryak:1977ut, HotQCD:2014kol}. In proton-proton (pp) collisions, the general assumption has historically been that temperature and density is too low for a QGP to form. However, at the LHC it was found that small system collisions exhibit several of the same characteristics as collisions of ions \cite{CMS:2010ifv, ALICE:2012eyl, ATLAS:2015hzw, ATLAS:2014qaj, ALICE:2016fzo}. In particular, the seminal measurement by the ALICE collaboration revealed that the strangeness production is enhanced as a function of event multiplicity, smoothly connecting the smallest multiplicities in \pp , with only a single gluon exchange, and the most central AA collisions \cite{ALICE:2016fzo, ALICE:2019avo}. 
\begin{figure} [hb]
\includegraphics[width=0.5\textwidth]{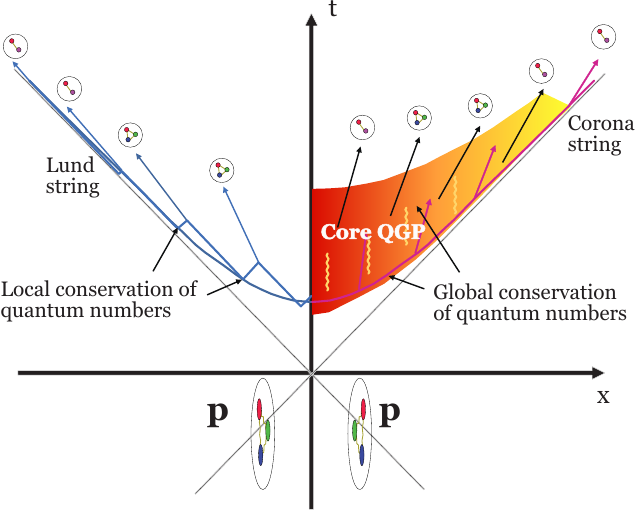}
\caption{Lightcone diagram sketch of the considered hadronization approaches, highlighting the main differences. The Lund string (left) conserves strangeness in individual string breaks. Hadrons are formed around a hyperbola in space-time. The core-corona (right) mixes a corona (magenta) string, which does not conserve strangeness locally, and a core QGP, which conserves strangeness globally over the full volume. As also indicated on the sketch, a QGP is much more long-lived than the string, which decays rapidly to hadrons after $\sqrt{\langle \tau^2 \rangle} \approx 1.4$ fm.}
\label{fig:lightcone-sketch}       % Give a unique label
\end{figure}
\begin{figure*}[ht]
\centering
\includegraphics[width=0.9\textwidth]{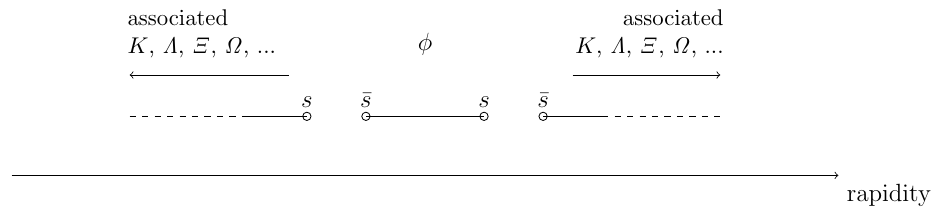}
\caption{When a $\phi$ meson is produced, it requires two neighbouring $s\bar{s}$ string breaks. This means that the neighbouring (associated) hadrons will include the leftover (anti-)strange quarks from the breakup.}
\label{fig:sketch-phi}
\end{figure*}

Two very different models have performed well in reproducing the increase of strangeness (see Supplemental Material \cite{SM}). Core-corona models \cite{Werner:2007bf} where the basic assumption is that droplets of QGP can form down to the smallest collision systems, and models of microscopic interactions of strings \cite{Sjostrand:1978dj, Andersson:1979ij, Andersson:1983jt} or clusters \cite{Webber:1983if} where fragmentation dynamics is modified by the presence of other strings or clusters, but no QGP is formed. In this paper we will focus on the \epos \cite{Werner:2023zvo} implementation of the core-corona model and the \pythia \cite{Bierlich:2022pfr} implementation of microscopic interactions, the rope-hadronisation model \cite{Bierlich:2014xba}. 

The current state of affairs, where the two types of models with qualitatively different hadronisation mechanisms and assumptions of QGP formation can explain the same signal, is unsatisfying. The key question is if there are observables for which the two models give very different predictions, that can be easily explained in terms of model differences. 
In this paper, we will develop a series of observables which satisfy these conditions based on the assumption that microscopically interacting strings preserve the strangeness quantum number locally (i.e.~in each breaking of a string), while a macroscopic model will necessarily only preserve it globally or in some volume of created QGP, see Fig.~\ref{fig:lightcone-sketch}.
\section{Two mechanisms for strangeness conservation}
The core-corona model \cite{Werner:2007bf} underlies hadronisation in the \epos event generator \cite{Werner:2023zvo}. 
In \epos, multiple scatterings hadronise, leading to the formation of prehadrons. Prehadrons close to the surface or escaping the bulk with a large transverse momentum will be treated as corona, i.e.~as in vacuum. The corona part hadronises using a string model. The \epos string does not break as an iterative cascade, where hadrons share break-up quarks. 
If the density of prehadrons is high, as e.g.~high multiplicity pp or AA collisions, they will be treated as core that thermalises and expands collectively, using a hydrodynamic treatment. For this study, the important part of the core evolution is the flavour assignment, which for small systems is performed using a microcanonical approach. In this approach, a volume $V$ with energy $E$ and net flavour content $Q = (n_u - n_{\bar{u}}, n_d - n_{\bar{d}}, n_s - n_{\bar{s}})$, decays into a configuration $\{h_1,...,h_n\}$ of hadrons given by the microcanonical partition function \cite{Werner:1995mx,Liu:2003cs}:
\begin{align}
  \Omega(\{h_1,...,h_n\}) = & \frac{V^n}{(2\pi)^{3n}} \prod_{i=1}^n g_i \prod_{\alpha \in S} \frac{1}{n_\alpha!} \nonumber \\
  & \int \prod^n_{i=1} d^3p_i \delta\left(E-\sum \varepsilon_i\right)  \\
  & \delta\left(\sum \vec{p}_i\right) \delta_{Q,\sum q_i}, \nonumber
\end{align}
where $\varepsilon_i = \sqrt{m^2_i - p^2_i}$ is the energy, $\vec{p}_i$ the 3-momentum, $n_\alpha$ the number of particles of species $\alpha$, $S$ the set of particle species considered, $g_i$ the degeneracy of particle $i$. The $\delta$-function at the end of the partition function ensures flavour conservation, with $q_i$ being the flavour composition of hadron $i$. The flavour conservation is over the full volume.

\begin{figure*}[t]
\includegraphics[width=0.49\textwidth]{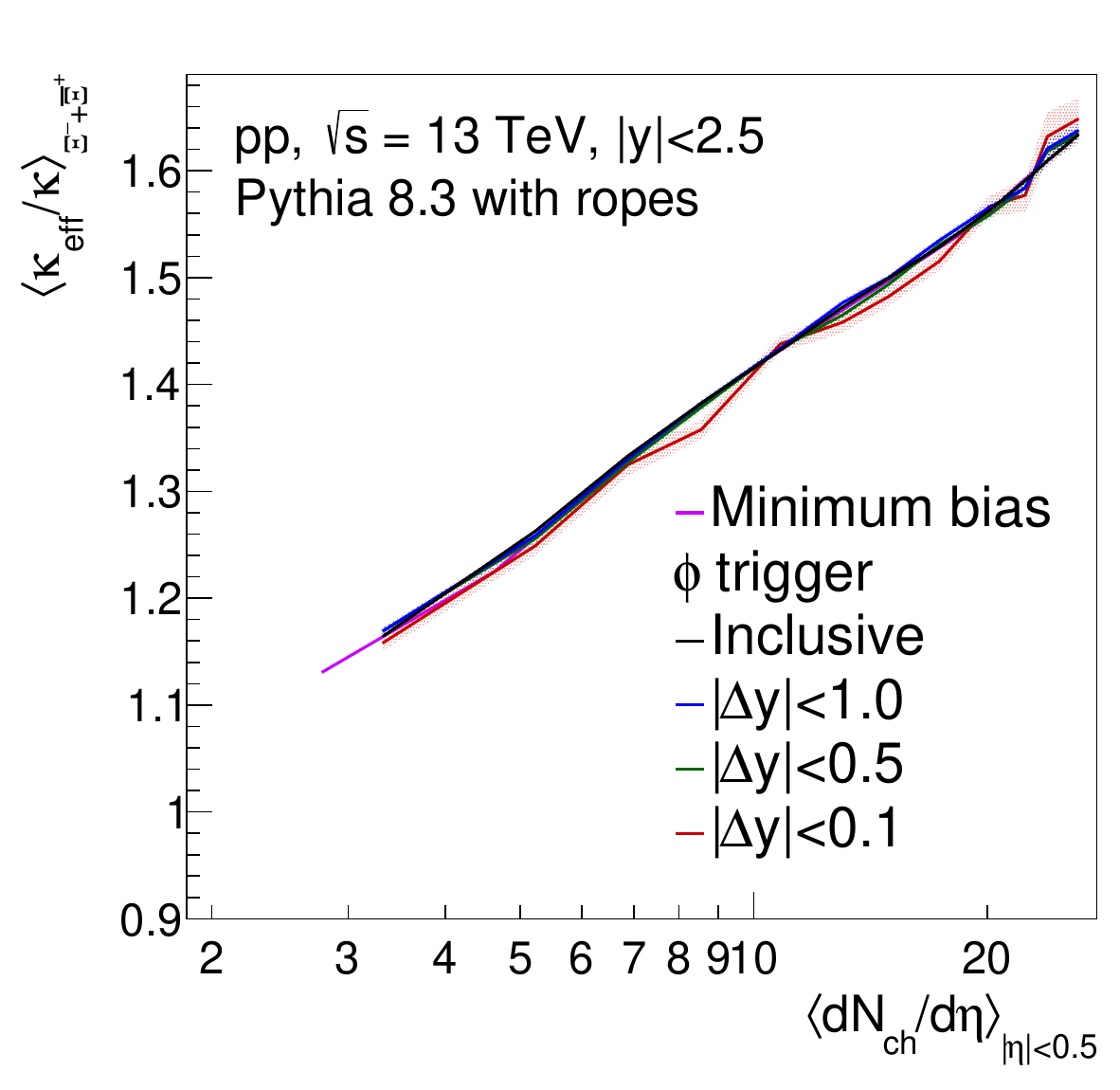}
\includegraphics[width=0.49\textwidth]{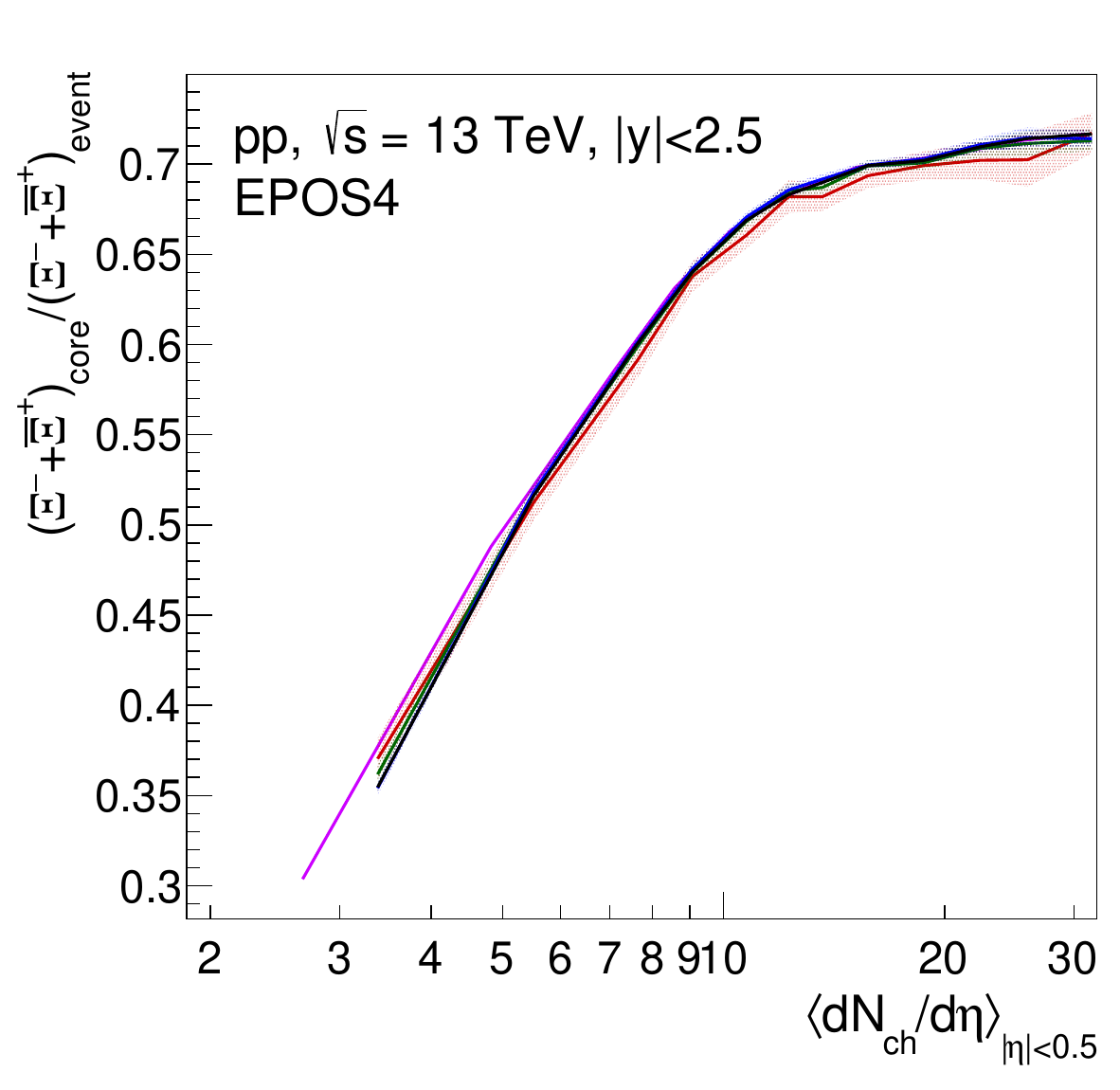}
\caption{Effective string tension (left) and core fraction (right) producing $(\Xi^-+\overline \Xi^+)$ in $|y|<2.5$ as a function of $\bigl\langle dN_{ch}/d\eta \bigr\rangle$.}
\label{fig:kappa-core}
\end{figure*}

The Lund string model is the model underlying hadronisation in the \pythia event generator \cite{Bierlich:2022pfr}. Several recent versions include the rope-hadronisation model \cite{Bierlich:2014xba, Bierlich:2016vgw, Bierlich:2017vhg, Bierlich:2024odg} relevant for this study. 
In \pythia, multiple 2 $\rightarrow$ 2 parton--parton scatterings radiate off additional quarks and gluons in the parton shower, to finally be hadronised. A $q\bar{q}$ pair will have a string spanned between them, with gluons acting as ``kinks'' on the string. One string hadronises to several hadrons, the amount determined by the initial energy of the $q\bar{q}$ pair, and approximately one hadron is produced per unit of rapidity per string. 
The hadron flavour is determined by additional $q\bar{q}$ pairs tunnelling out of the vacuum
when the string breaks (plus the flavour of the two additional end-point quarks). In string breaks, strange quarks are suppressed with respect to up or down by a factor:
\begin{equation}
        \rho = \exp\left(-\frac{\pi(m^2_s - m^2_u)}{\kappa}\right),
\label{eq:rho}\end{equation}
where $m_s$ and $m_u$ are the strange and light quark masses respectively, and $\kappa$ is the string tension, which has a numerical value around 1 GeV/fm. In practice, due to uncertainty 
about which value to use for quark masses, $\rho$ is treated as a free parameter, estimated in $e^+e^-$ collisions at the Z pole, i.e.~a single string system. The current default value is $\rho = 0.217~$\cite{Bierlich:2022pfr}.
When several strings overlap with each other, junction topologies which carry an intrinsic baryon number can be dynamically formed \cite{Sjostrand:2002ip, Christiansen:2015yqa}, and the effective string tension $\kappa_{\text{eff}}$ in the string breaks increases. Starting from the value estimated in $e^+e^-$, the parameter $\rho$ in eq.~\ref{eq:rho} will increase as the effective string tension increases \cite{Bierlich:2014xba}, giving rise to an enhancement of strange hadrons being formed in high-density environments.

In pp collisions, most strings are aligned with the rapidity axis of the laboratory system. The $\phi$ meson has a special role, when it is produced requiring two $s\bar{s}$ string breaks the leftover (anti-)strange quarks must find their home in hadrons produced in neighbouring rank of the same string, i.e.~close in rapidity to the $\phi$ (see Fig.~\ref{fig:sketch-phi}). 
This is a unique prediction of the Lund string model. If the string degrees of freedom would vanish, because they melt into a QGP, the structure carrying the correlations in rapidity would not be observed.
The explanation given above, holds for a single string. In pp collisions, many strings are produced in the same rapidity interval, and therefore one cannot be certain that a given pair of neighbouring particles are produced with string breaks in common, even though they may share quark pairs. In balance functions \cite{Bass:2000az, Pratt:2011bc, ALICE:2023asw}, this is handled by subtraction procedures. Here, however, we are aiming at a method that allows us to highlight directly the qualitative differences arising between the models. 

The main effect of applying a $\phi$ trigger will be the emergence of different types of correlations among produced hadrons. For the rope-hadronisation model, the effective string tension used to produce particles of a given species is studied with and without applying a $\phi$ trigger to test if the rapidity regions are biased to have a larger string tension. In Fig.~\ref{fig:kappa-core} (left) results are shown for $\Xi$ ($K_S^0$, $\Lambda$ and $\Omega$ yield to similar results and are discussed in the Supplemental Material \cite{SM}). In minimum bias results (magenta) no $\phi$ trigger has been applied, in the inclusive sample (black) a $\phi$ is required to be anywhere in the event within the selected acceptance ($|y| < 2.5$). The three selections of $|\Delta y|$ signifying that a $\phi$ meson is required to be within that rapidity range from the particle of interest. 
There is an enhancement in $\kappa_{eff}/\kappa$ as a function of multiplicity irrespective of the presence of a $\phi$, without saturating at high multiplicities. Therefore, the presence of a $\phi$ is not indicative of a larger local string tension than the one already indicated by the presence of a $\Xi$.

In Fig.~\ref{fig:kappa-core} (right) a similar study is repeated for the core-corona model implemented in \epos to test if a larger fraction of particles are produced from the core, when a rapidity window around the $\phi$ is selected.  The quantity of interest in this case is the core fraction, defined as the amount of particles of a given species coming from the core, divided by the total amount of that species in the event. The conclusions are similar. The presence of the $\phi$ does not enhance the core fraction for the particle of interest, although in general the strange baryons are produced more copiously in the core. A noteworthy difference between the two models is that the core fraction saturates at high multiplicities, where the effective string tension does not.

The combined results of Fig.~\ref{fig:kappa-core} confirm that the observables presented in the next section will not bias towards a local enhanced string tension or core fraction, but are well suited for studying particle correlations.

\begin{figure*}[t]
\includegraphics[width=0.49\textwidth]{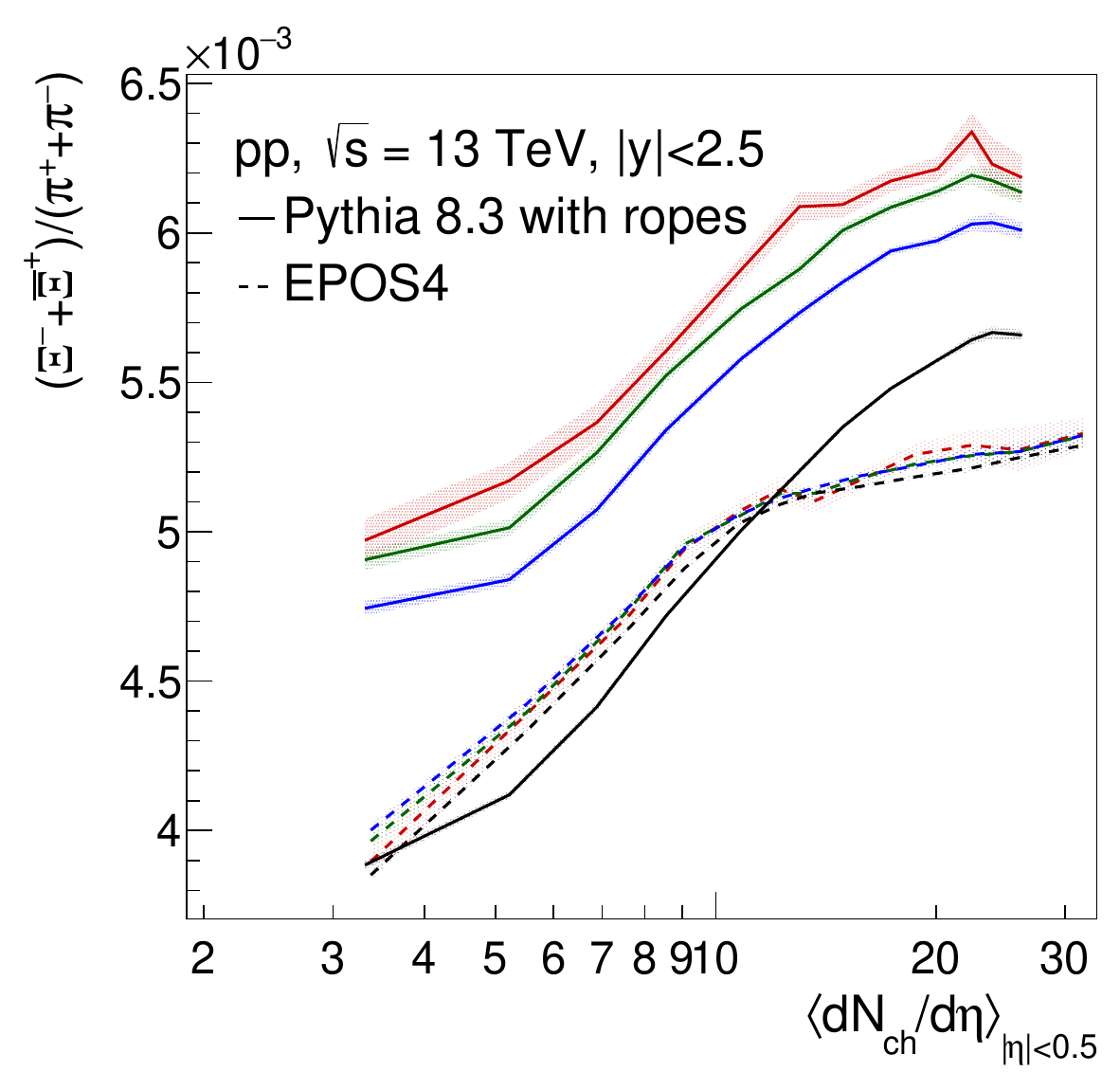}
\includegraphics[width=0.49\textwidth]{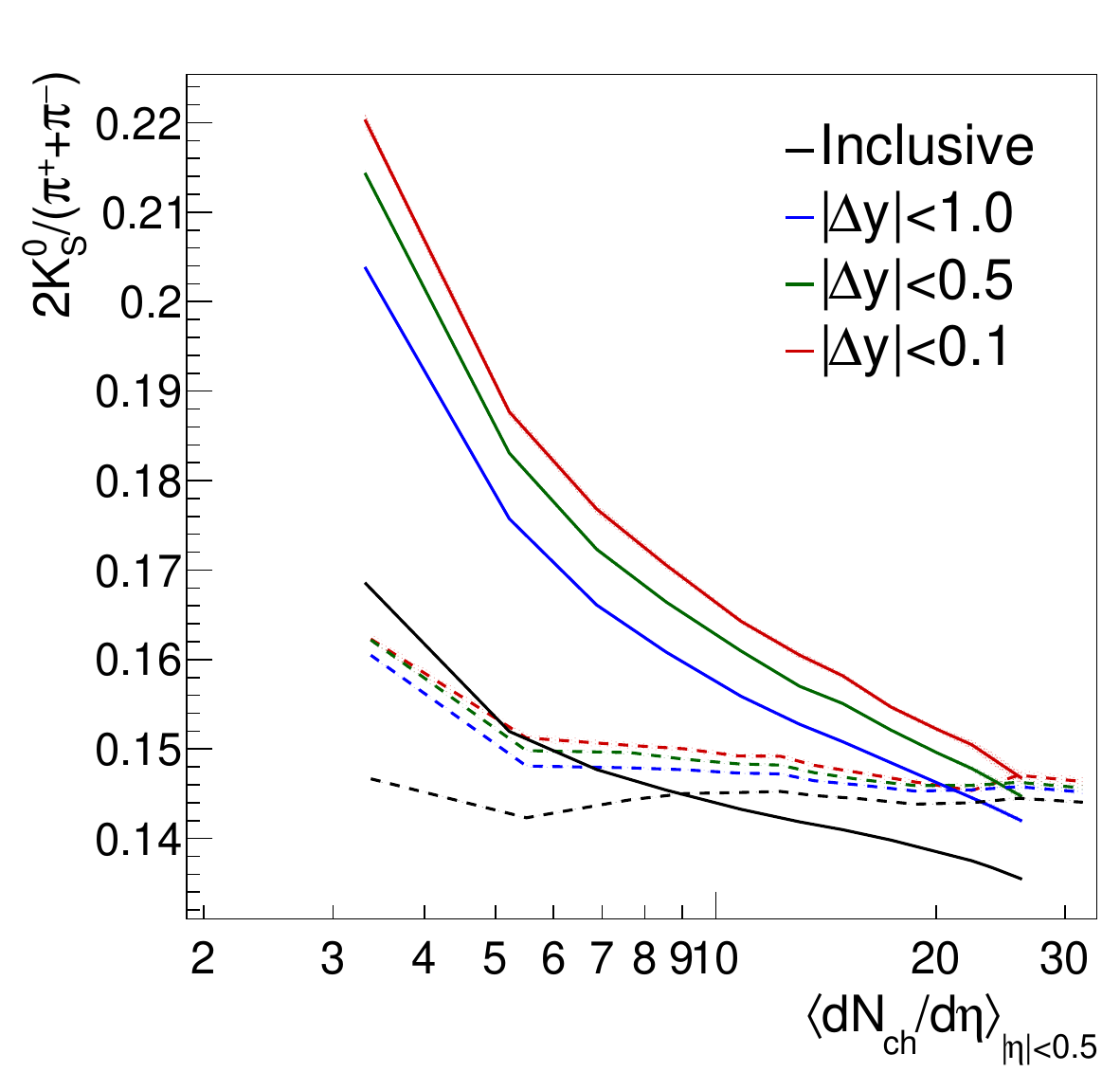}
\caption{Yield ratios of $(\Xi^-+\overline \Xi^+)/(\pi^++\pi^-)$ (left) and $2K_S^0/(\pi^++\pi^-)$ (right) as a function of $\bigl\langle dN_{ch}/d\eta \bigr\rangle$ in $|\eta|<0.5$.}
\label{fig:enhance-xi-k}
\end{figure*}

\section{Results with final state particles}

The results at hadron level will be now discussed. The suggested observables to distinguish between the two types of models are particle ratios in rapidity intervals ($\Delta y$) around the triggered $\phi$ meson. 
The inclusive results (black) indicate that the whole $|y| < 2.5$ is considered with the presence of a $\phi$ meson in the event. In Fig~\ref{fig:enhance-xi-k} ratios of (multi-)strange hadron yields to pions are shown, as a function of charged-particle multiplicity in the rapidity range, $|y|<2.5$. The coloured lines represent the different rapidity separations between the $\phi$ meson and the strange hadron of interest, in red $|\Delta y|<0.1$,  in green $|\Delta y|<0.5$, and in blue $|\Delta y|<1.0$. \pythia is shown with fully drawn lines, \epos with dashed lines.

In Fig.~\ref{fig:enhance-xi-k} (left) results for $\Xi$ are shown. As the core-corona model is insensitive to local flavour conservation at the level of string breaks, the \epos curves collapse into one single curve. \pythia, on the other hand, observes local flavour conservation at the level of string breaks, and a marked enhancement is clearly seen moving from the inclusive sample, and closer to the $\phi$. As one would expect, the effect quickly saturates -- the curves for the three $|\Delta y|$ selections are very close together -- as hadrons are produced with a separation of one unit of rapidity between them on average.
All curves, for both models, show some degree of saturation at high multiplicities, although this feature is much more noticeable for \epos. As shown in Fig.~\ref{fig:kappa-core} the core fraction itself saturates while the effective string tension does not.

The results for $\Lambda$ are qualitatively similar to those presented in Fig~\ref{fig:enhance-xi-k} (left) and are discussed in the Supplemental Material \cite{SM} together with the results for $\Omega$, characterised by a di-quark followed by an additional $s$-quark production mechanism implemented in \pythia \cite{Andersson:1981ce}.

In Fig. \ref{fig:enhance-xi-k} (right) results for $K^0_S$ are shown. The $K^0_S$ yields with respect to pions are remarkably different in the two models. The \pythia curves decrease towards high multiplicities, opposite to what one would naively expect from QGP strangeness enhancement and from what one observes in the Minimum Bias figure contained in the Supplemental Material \cite{SM}. The explanation lies in local conservation of strangeness at the level of string breaks. At low multiplicity, there are no baryon junctions. If the presence of a $\phi$ meson is requested the additional strange quark will allow the production of a $K^0_S$ (see Fig.~\ref{fig:sketch-phi}). In the extreme case of the lowest multiplicity bin with just a few charged particles produced per unit of rapidity, two of those charged particles are likely to be pions from the $\phi$ decay. This means that it is much more probable to have a $K^0_S$ produced on each side of the $\phi$, rather than producing a heavily suppressed strange or multi-strange baryon. 
The \epos curves are again almost collapsed to a single curve, with a slight bend upwards at low multiplicities that might be attributed to the extreme low multiplicity that forces the presence of a handful of charged particles only, a $\phi$ and a $K^0_S$ in the event.  
Hadron to pion ratios as double ratios are shown and discussed in the Supplemental Material \cite{SM}. 

\section{Considerations and outlook}

In conclusion, the seminal strangeness enhancement results of the ALICE experiment started a whole new line of research linking flavour enhancement in small and large systems. The apparent ability of microscopic string based models and models based on QGP formation to both describe the data equally well has spurred a large interest in finding observables which can discriminate between the two types of models.
In this paper we have developed observables for which the underlying physics differences of the models are highlighted, and that produce significantly different and measurable results for strange-hadron production in events with a $\phi$ meson.

Two main observations were made. Firstly, \pythia with ropes predicts a clear layering of the strange hadron to pion ratios as a function of the charged particle multiplicity which is not observed in \epos. This is expected to originate from the Lund strings and rope-hadronisation model that imply local conservation of strangeness. Secondly,  the $K^0_S/\pi$ ratio is decreasing with multiplicity for \pythia with ropes, remaining rather flat for \epos. This is in disagreement with the results both from data and models in the case of MB $K^0_S/\pi$ ratio, where no $\phi$ trigger is requested. 

Both these observations will have to be tested in data. Run 3 and 4 at LHC open the possibility for such measurements which are statistics eager. Indeed, the amount of data that will be collected will be one to two orders of magnitude larger with respect to the amount of data collected in Run 1 and 2. Another improvement will come from the detector upgrades which will allow to reconstruct particle yields at lower transverse momentum and with higher momentum resolution. Innovative reconstruction algorithms will permit to directly track strange particles, due to vicinity of inner trackers to the beam pipe. Thanks to these improvements it will be possible to detach the different strange-hadron production mechanisms in vacuum and dense environments.

\begin{acknowledgements}
Support from the Knut and Alice Wallenberg foundation contract number 2017.0036 (CB and SC), and Vetenskapsrådet contracts 2016-05996 and 2023-04316 (CB) is gratefully acknowledged.
\end{acknowledgements}

\bibliographystyle{spmpsci}      % mathematics and physical sciences
\bibliography{bibliography}   % name your BibTeX data base

\begin{thebibliography}{10}
\providecommand{\url}[1]{{#1}}
\providecommand{\urlprefix}{URL }
\expandafter\ifx\csname urlstyle\endcsname\relax
  \providecommand{\doi}[1]{DOI~\discretionary{}{}{}#1}\else
  \providecommand{\doi}{DOI~\discretionary{}{}{}\begingroup \urlstyle{rm}\Url}\fi

\bibitem{ATLAS:2014qaj}
Aad, G., et~al.: {Measurement of long-range pseudorapidity correlations and azimuthal harmonics in $\sqrt{s_{NN}}=5.02$ TeV proton-lead collisions with the ATLAS detector}.
\newblock Phys. Rev. C \textbf{90}(4), 044,906 (2014).
\newblock \doi{10.1103/PhysRevC.90.044906}

\bibitem{ATLAS:2015hzw}
Aad, G., et~al.: {Observation of Long-Range Elliptic Azimuthal Anisotropies in $\sqrt{s}=$13 and 2.76 TeV $pp$ Collisions with the ATLAS Detector}.
\newblock Phys. Rev. Lett. \textbf{116}(17), 172,301 (2016).
\newblock \doi{10.1103/PhysRevLett.116.172301}

\bibitem{ALICE:2012eyl}
Abelev, B., et~al.: {Long-range angular correlations on the near and away side in $p$-Pb collisions at $\sqrt{s_{NN}}=5.02$ TeV}.
\newblock Phys. Lett. B \textbf{719}, 29--41 (2013).
\newblock \doi{10.1016/j.physletb.2013.01.012}

\bibitem{ALICE:2019avo}
Acharya, S., et~al.: {Multiplicity dependence of (multi-)strange hadron production in proton-proton collisions at $\sqrt{s}$ = 13 TeV}.
\newblock Eur. Phys. J. C \textbf{80}(2), 167 (2020).
\newblock \doi{10.1140/epjc/s10052-020-7673-8}

\bibitem{ALICE:2023asw}
Acharya, S., et~al.: {Studying strangeness and baryon production mechanisms through angular correlations between charged $\Xi$ baryons and identified hadrons in pp collisions at $\sqrt{s}$ = 13 TeV}  (2023)

\bibitem{ALICE:2016fzo}
Adam, J., et~al.: {Enhanced production of multi-strange hadrons in high-multiplicity proton-proton collisions}.
\newblock Nature Phys. \textbf{13}, 535--539 (2017).
\newblock \doi{10.1038/nphys4111}

\bibitem{Andersson:1979ij}
Andersson, B., Gustafson, G.: {Semiclassical Models for Gluon Jets and Leptoproduction Based on the Massless Relativistic String}.
\newblock Z. Phys. C \textbf{3}, 223 (1980).
\newblock \doi{10.1007/BF01577421}

\bibitem{Andersson:1981ce}
Andersson, B., Gustafson, G., Sjostrand, T.: {A Model for Baryon Production in Quark and Gluon Jets}.
\newblock Nucl. Phys. B \textbf{197}, 45--54 (1982).
\newblock \doi{10.1016/0550-3213(82)90153-5}

\bibitem{Andersson:1983jt}
Andersson, B., Gustafson, G., Soderberg, B.: {A General Model for Jet Fragmentation}.
\newblock Z. Phys. C \textbf{20}, 317 (1983).
\newblock \doi{10.1007/BF01407824}

\bibitem{Bass:2000az}
Bass, S.A., Danielewicz, P., Pratt, S.: {Clocking hadronization in relativistic heavy ion collisions with balance functions}.
\newblock Phys. Rev. Lett. \textbf{85}, 2689--2692 (2000).
\newblock \doi{10.1103/PhysRevLett.85.2689}

\bibitem{HotQCD:2014kol}
Bazavov, A., et~al.: {Equation of state in ( 2+1 )-flavor QCD}.
\newblock Phys. Rev. D \textbf{90}, 094,503 (2014).
\newblock \doi{10.1103/PhysRevD.90.094503}

\bibitem{Bierlich:2024odg}
Bierlich, C.: {String Interactions as a Source of Collective Behaviour}.
\newblock Universe \textbf{10}(1), 46 (2024).
\newblock \doi{10.3390/universe10010046}

\bibitem{SM}
Bierlich, C., Cannito, S., Zaccolo, V.: {A test of strangeness quantum number conservation in proton-proton collisions}

\bibitem{Bierlich:2016vgw}
Bierlich, C., Gustafson, G., L\"onnblad, L.: {A shoving model for collectivity in hadronic collisions}  (2016)

\bibitem{Bierlich:2017vhg}
Bierlich, C., Gustafson, G., L\"onnblad, L.: {Collectivity without plasma in hadronic collisions}.
\newblock Phys. Lett. B \textbf{779}, 58--63 (2018).
\newblock \doi{10.1016/j.physletb.2018.01.069}

\bibitem{Bierlich:2014xba}
Bierlich, C., Gustafson, G., L\"onnblad, L., Tarasov, A.: {Effects of Overlapping Strings in pp Collisions}.
\newblock JHEP \textbf{03}, 148 (2015).
\newblock \doi{10.1007/JHEP03(2015)148}

\bibitem{Bierlich:2022pfr}
Bierlich, C., et~al.: {A comprehensive guide to the physics and usage of PYTHIA 8.3}.
\newblock SciPost Phys. Codeb. \textbf{2022}, 8 (2022).
\newblock \doi{10.21468/SciPostPhysCodeb.8}

\bibitem{Cabibbo:1975ig}
Cabibbo, N., Parisi, G.: {Exponential Hadronic Spectrum and Quark Liberation}.
\newblock Phys. Lett. B \textbf{59}, 67--69 (1975).
\newblock \doi{10.1016/0370-2693(75)90158-6}

\bibitem{Christiansen:2015yqa}
Christiansen, J.R., Skands, P.Z.: {String Formation Beyond Leading Colour}.
\newblock JHEP \textbf{08}, 003 (2015).
\newblock \doi{10.1007/JHEP08(2015)003}

\bibitem{CMS:2010ifv}
Khachatryan, V., et~al.: {Observation of Long-Range Near-Side Angular Correlations in Proton-Proton Collisions at the LHC}.
\newblock JHEP \textbf{09}, 091 (2010).
\newblock \doi{10.1007/JHEP09(2010)091}

\bibitem{Liu:2003cs}
Liu, F.M., Werner, K., Aichelin, J.: {Comparison of microcanonical and canonical hadronization}.
\newblock Phys. Rev. C \textbf{68}, 024,905 (2003).
\newblock \doi{10.1103/PhysRevC.68.024905}

\bibitem{Pratt:2011bc}
Pratt, S.: {General Charge Balance Functions, A Tool for Studying the Chemical Evolution of the Quark-Gluon Plasma}.
\newblock Phys. Rev. C \textbf{85}, 014,904 (2012).
\newblock \doi{10.1103/PhysRevC.85.014904}

\bibitem{Shuryak:1977ut}
Shuryak, E.V.: {Theory of Hadronic Plasma}.
\newblock Sov. Phys. JETP \textbf{47}, 212--219 (1978)

\bibitem{Sjostrand:2002ip}
Sjostrand, T., Skands, P.Z.: {Baryon number violation and string topologies}.
\newblock Nucl. Phys. B \textbf{659}, 243 (2003).
\newblock \doi{10.1016/S0550-3213(03)00193-7}

\bibitem{Sjostrand:1978dj}
Sjostrand, T., Soderberg, B.: {A MONTE CARLO PROGRAM FOR QUARK JET GENERATION}  (1978)

\bibitem{Webber:1983if}
Webber, B.R.: {A QCD Model for Jet Fragmentation Including Soft Gluon Interference}.
\newblock Nucl. Phys. B \textbf{238}, 492--528 (1984).
\newblock \doi{10.1016/0550-3213(84)90333-X}

\bibitem{Werner:2007bf}
Werner, K.: {Core-corona separation in ultra-relativistic heavy ion collisions}.
\newblock Phys. Rev. Lett. \textbf{98}, 152,301 (2007).
\newblock \doi{10.1103/PhysRevLett.98.152301}

\bibitem{Werner:2023zvo}
Werner, K.: {Revealing a deep connection between factorization and saturation: New insight into modeling high-energy proton-proton and nucleus-nucleus scattering in the EPOS4 framework}.
\newblock Phys. Rev. C \textbf{108}(6), 064,903 (2023).
\newblock \doi{10.1103/PhysRevC.108.064903}

\bibitem{Werner:1995mx}
Werner, K., Aichelin, J.: {Microcanonical treatment of hadronizing the quark - gluon plasma}.
\newblock Phys. Rev. C \textbf{52}, 1584--1603 (1995).
\newblock \doi{10.1103/PhysRevC.52.1584}

\end{thebibliography}

\end{document}

\title{Supplemental material \\
A test of strangeness quantum number conservation \\ in proton-proton collisions}

\author{Christian Bierlich\thanksref{e1,addr1}
        \and
        Stefano Cannito\thanksref{e2,addr2} 
        \and 
        Valentina Zaccolo\thanksref{e3,addr2} 
}

\thankstext{e1}{e-mail: christian.bierlich@fysik.lu.se}
\thankstext{e2}{e-mail: stefano.cannito@studenti.units.it}
\thankstext{e3}{e-mail: valentina.zaccolo@units.it}

\institute{Department of Physics, Lund University, Box 118, Lund, SE-221 00, Sweden\label{addr1}
\and
Department of Physics, University of Trieste, Via Alfonso Valerio 2, Trieste, 34127, Italy\label{addr2}
}

\date{Received: date / Accepted: date}

\maketitle

\begin{abstract}
This Supplemental Material contains additional figures for strange-hadron yields and the settings of \epos and \pythia utilised for the generation.
\end{abstract}

\section{Introduction}
\begin{figure}[hb]
\includegraphics[width=0.5\textwidth]{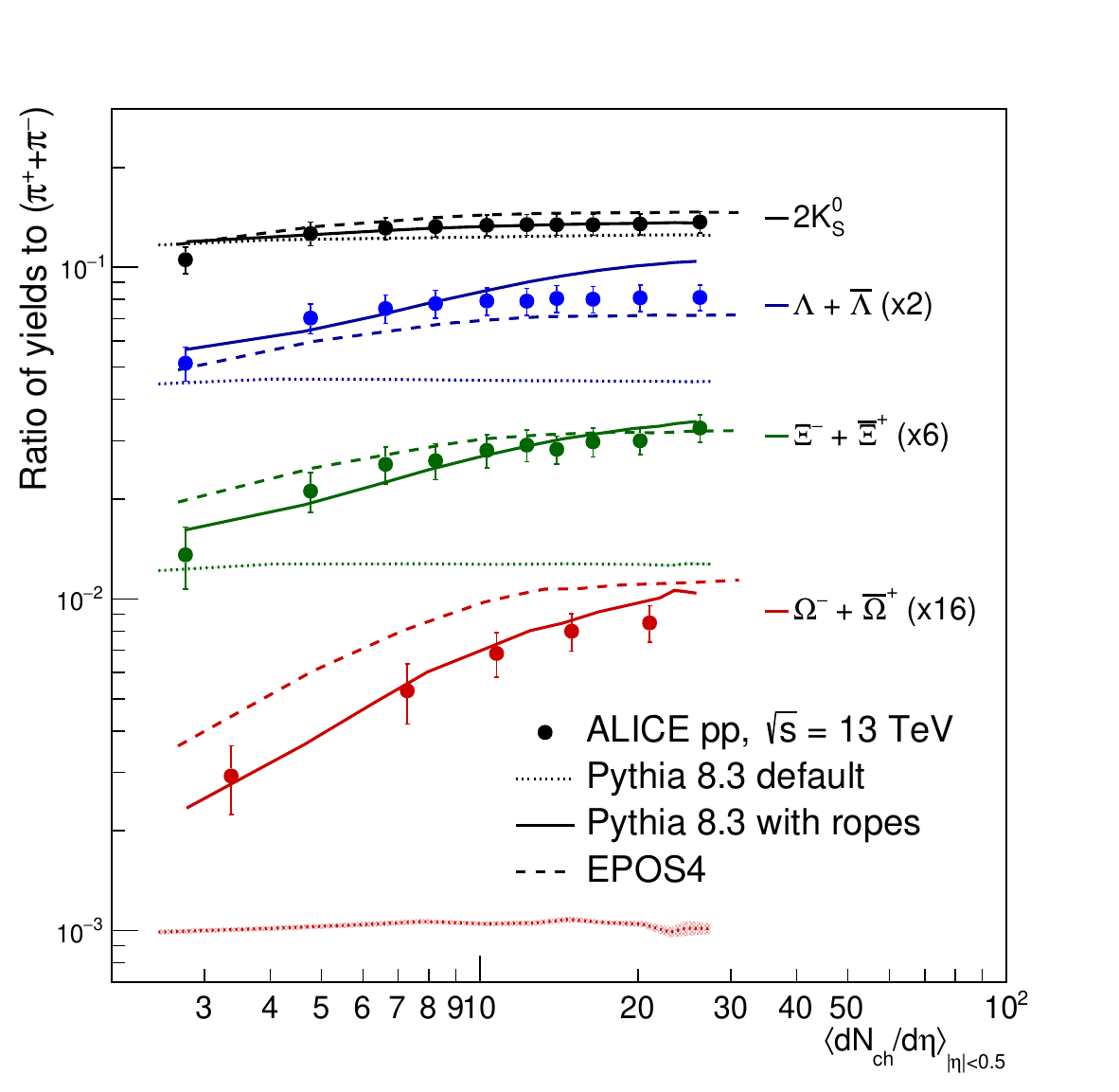}
\caption{Yield ratios of \(K_S^0,\ \Lambda+\overline\Lambda,\ \Xi^-+\overline\Xi^+\) and \(\Omega^-+\overline\Omega^+\) to \(\pi^++\pi^-\) in $|y|<0.5$ as a function of \(\bigl\langle dN_{ch}/d\eta \bigr\rangle_{|\eta|<0.5}\). Calculations from \pythia and \epos compared to ALICE results \cite{ALICE:2019avo} to show that both models do well in describing overall features of strangeness enhancement.}
\label{fig:strangeness-enhancement}
\end{figure}

As highlighted in the main publication, the two different models \epos and \pythia are able to reproduce the overall strangeness enhancement results by ALICE \cite{ALICE:2019avo} as it can be seen in Fig.~\ref{fig:strangeness-enhancement}.

\section{Results for $\Lambda$ and $\Omega$}
\begin{figure*}[ht]
\includegraphics[width=0.49\textwidth]{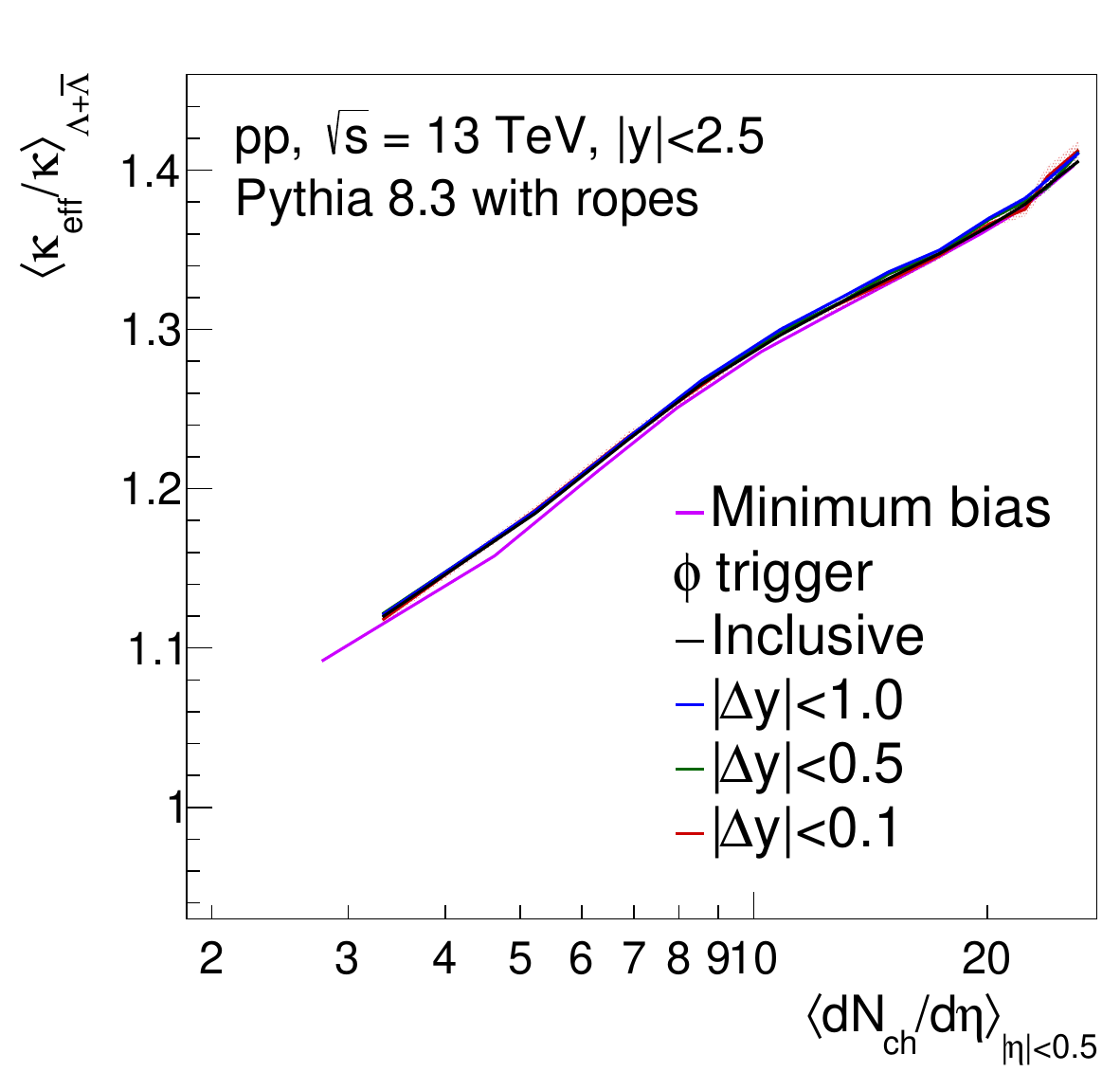}
\includegraphics[width=0.49\textwidth]{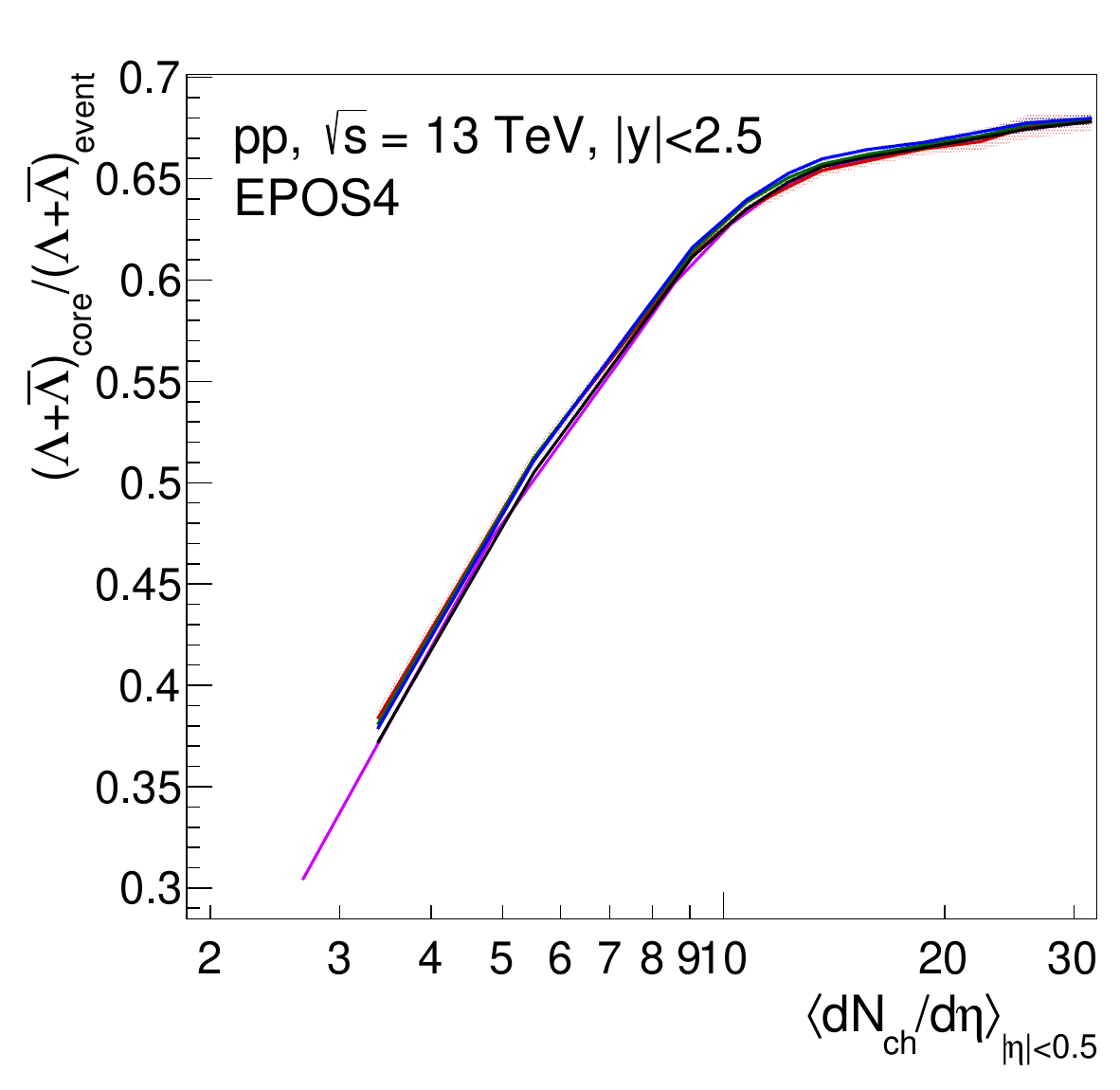}
\caption{Effective string tension (left) and core fraction (right) producing $(\Lambda^-+\overline \Lambda^+)$ in $|y|<2.5$ as a function of $\bigl\langle dN_{ch}/d\eta \bigr\rangle$.}
\label{fig:kappa-core}
\end{figure*}

Before presenting the hadron-level results in the publication, we presented the effective string tension in \pythia and the core fraction used in \epos to produce the $\Xi$ hadron, with and without applying a $\phi$ trigger. In Fig.~\ref{fig:kappa-core} results are shown for $\Lambda$ ($K_S^0$ and $\Omega$ yield to similar results). Likewise the case for $\Xi$, the presence of a $\phi$ is not indicative of a larger local string tension or core fraction than the one already indicated by the presence of a $\Lambda$.

\begin{figure*}[htbp]
\includegraphics[width=0.49\textwidth]{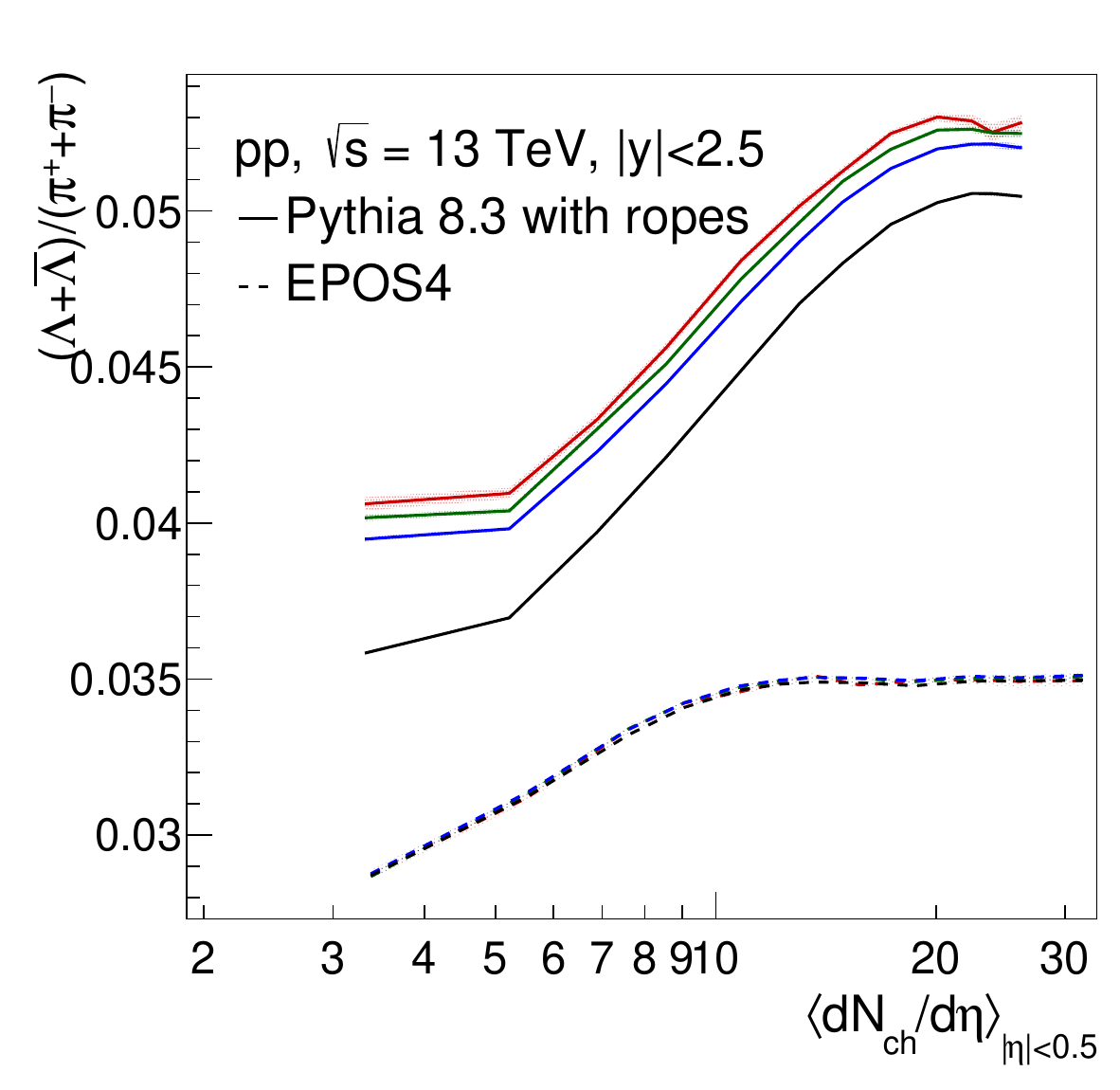}
\includegraphics[width=0.49\textwidth]{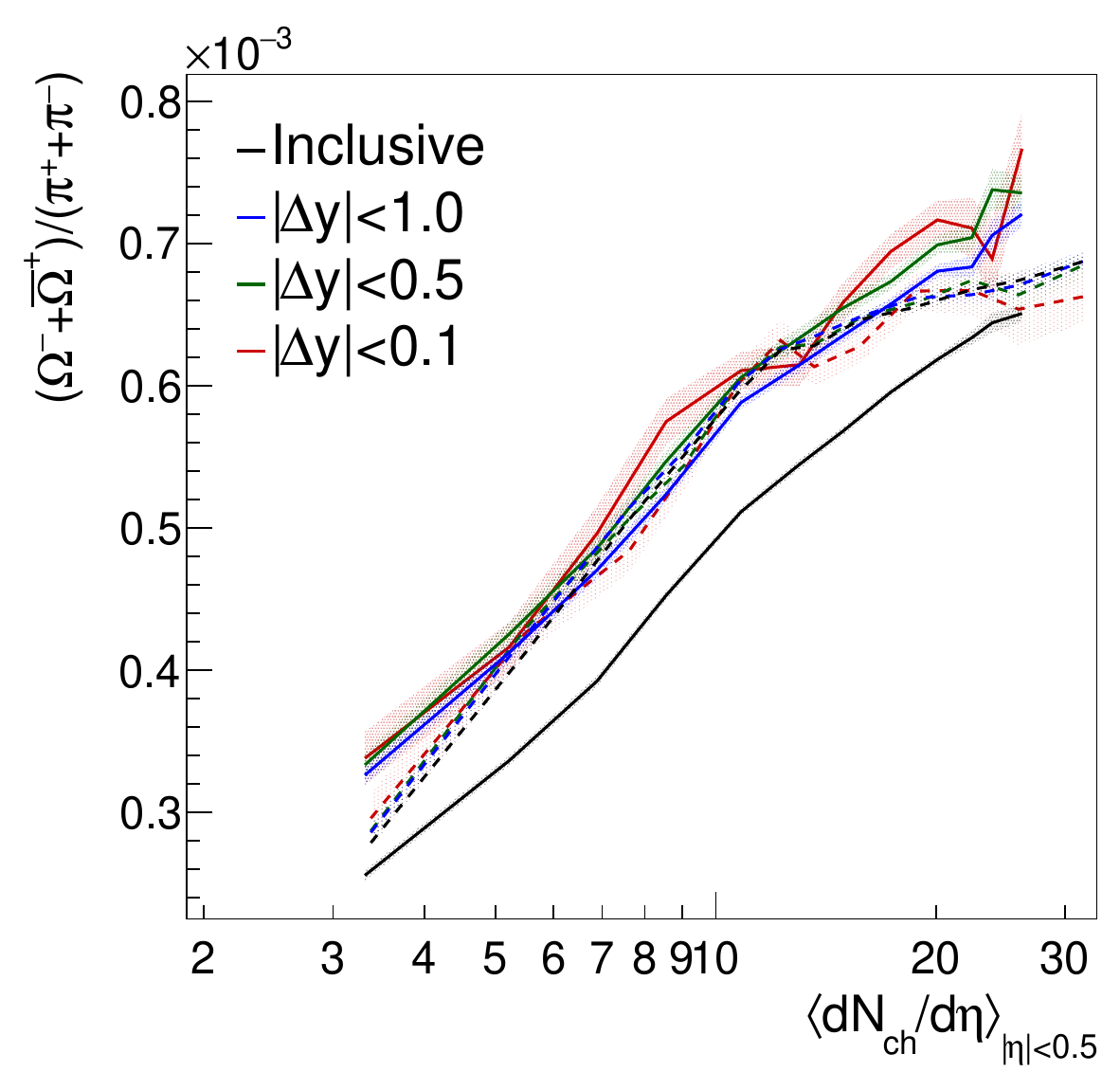}
\caption{Yield ratios of $(\Lambda^-+\overline \Lambda^+)/(\pi^++\pi^-)$ (left) and $(\Omega^-+\overline \Omega^+)/(\pi^++\pi^-)$ (right) as a function of $\bigl\langle dN_{ch}/d\eta \bigr\rangle$ in $|\eta|<0.5$.}
\label{fig:enhance-lambda-omega}
\end{figure*}

The results of particle ratios to pions in rapidity intervals ($\Delta y$) around the triggered $\phi$ meson have been presented in the main publication for $\Xi$ and $K^0_S$. In Fig. \ref{fig:enhance-lambda-omega}, we show results for $\Lambda$ (left) and $\Omega$ (right). The inclusive results (black) indicate that the whole $|y| < 2.5$ is considered, with the presence of a $\phi$-meson in the event.  The coloured lines represent the different rapidity separations between the $\phi$ meson and the strange hadron of interest, in red $|\Delta y|<0.1$,  in green $|\Delta y|<0.5$, and in blue $|\Delta y|<1.0$. \pythia is shown with fully drawn lines, \epos with dashed lines.

The results for $\Lambda$ (left) are qualitatively similar to those presented for the $\Xi$. The \epos curves collapse into one single curve. Instead a marked enhancement is clearly seen moving from the inclusive sample and closer to the $\phi$ for \pythia with ropes. 

All curves for $\Omega$ in Fig. \ref{fig:enhance-lambda-omega} (right) are more or less collapsed into a single curve, even though additional strange quarks are, in the case of \pythia, still present from the $\phi$ production. 

In \pythia, the only way to produce the $\Omega$ baryon is by production of an $ss_1$ diquark (subscript indicates spin) followed by an $s$-quark \cite{Andersson:1981ce}. The enhancement observed indicates that production of the additional strange quark is not a limiting factor for $\Omega$ production once an $ss_1$ diquark is produced. There are no octet baryons with quark content $sss$, therefore a SU(6) spin$\times$flavour Clebsch-Gordan weight of zero is assigned to the $\Omega$ for the non-existing octet. This means that for $\Omega$ in the decuplet the weight will be of unity. 
If on the contrary a $u$ or a $d$ quark was produced in combination with the $ss_1$ diquark, like for the $\Xi$, the corresponding weights with the same normalisation amount to 1/6 in the octet case and 1/3 in the decuplet case. Since these two weights do not sum to unity, but rather to 1/2, it means that one effectively rejects half the $u$ or $d$ string breaks when they appear next to an $ss_1$ diquark. Consequently, the regular strangeness suppression factor no longer acts as a suppressor for $\Omega$-production.

\section{Double ratios}
\begin{figure*}[htbp]
\centering
\includegraphics[width=0.49\textwidth]{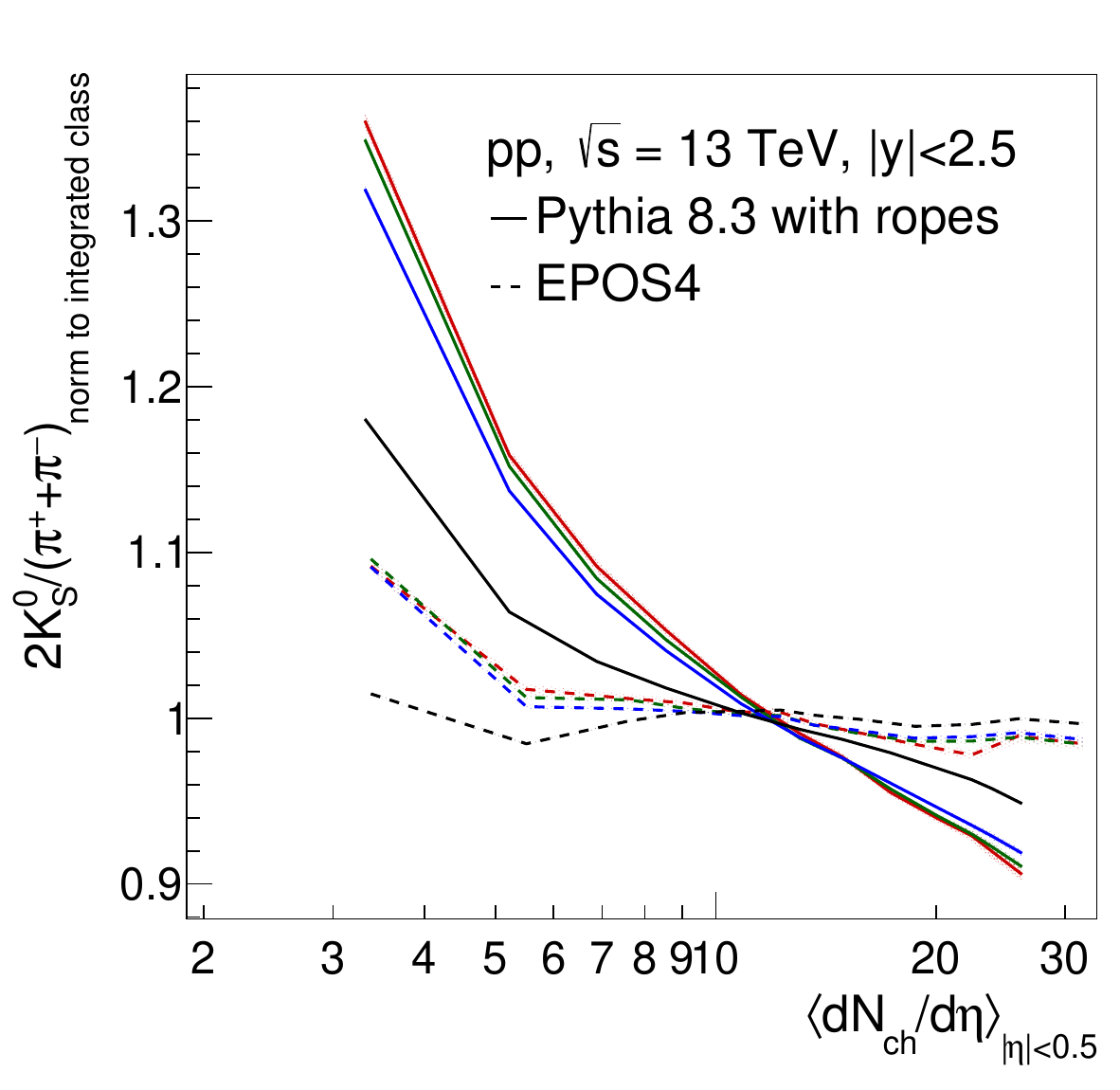}
\includegraphics[width=0.49\textwidth]{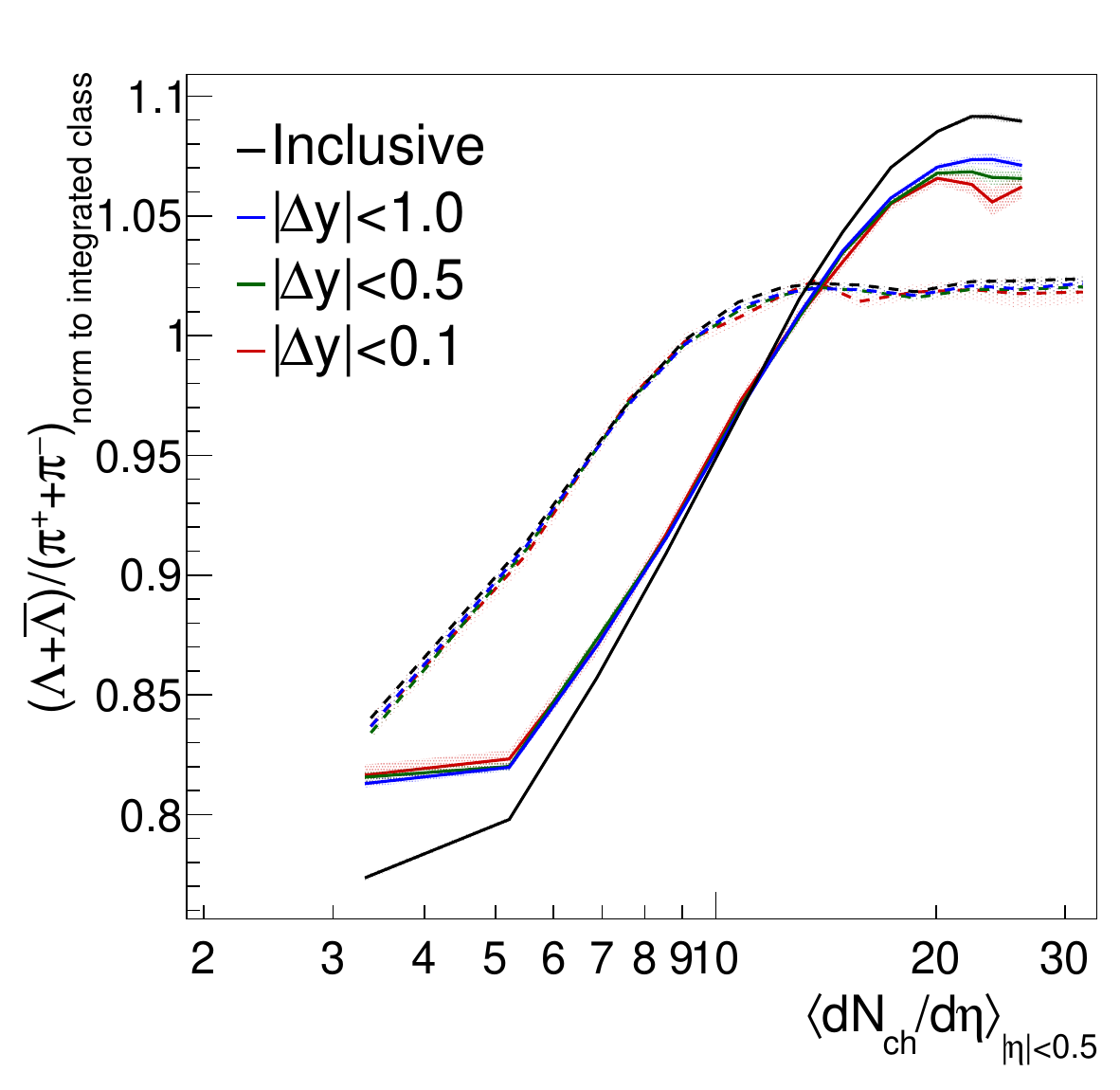}
\includegraphics[width=0.49\textwidth]{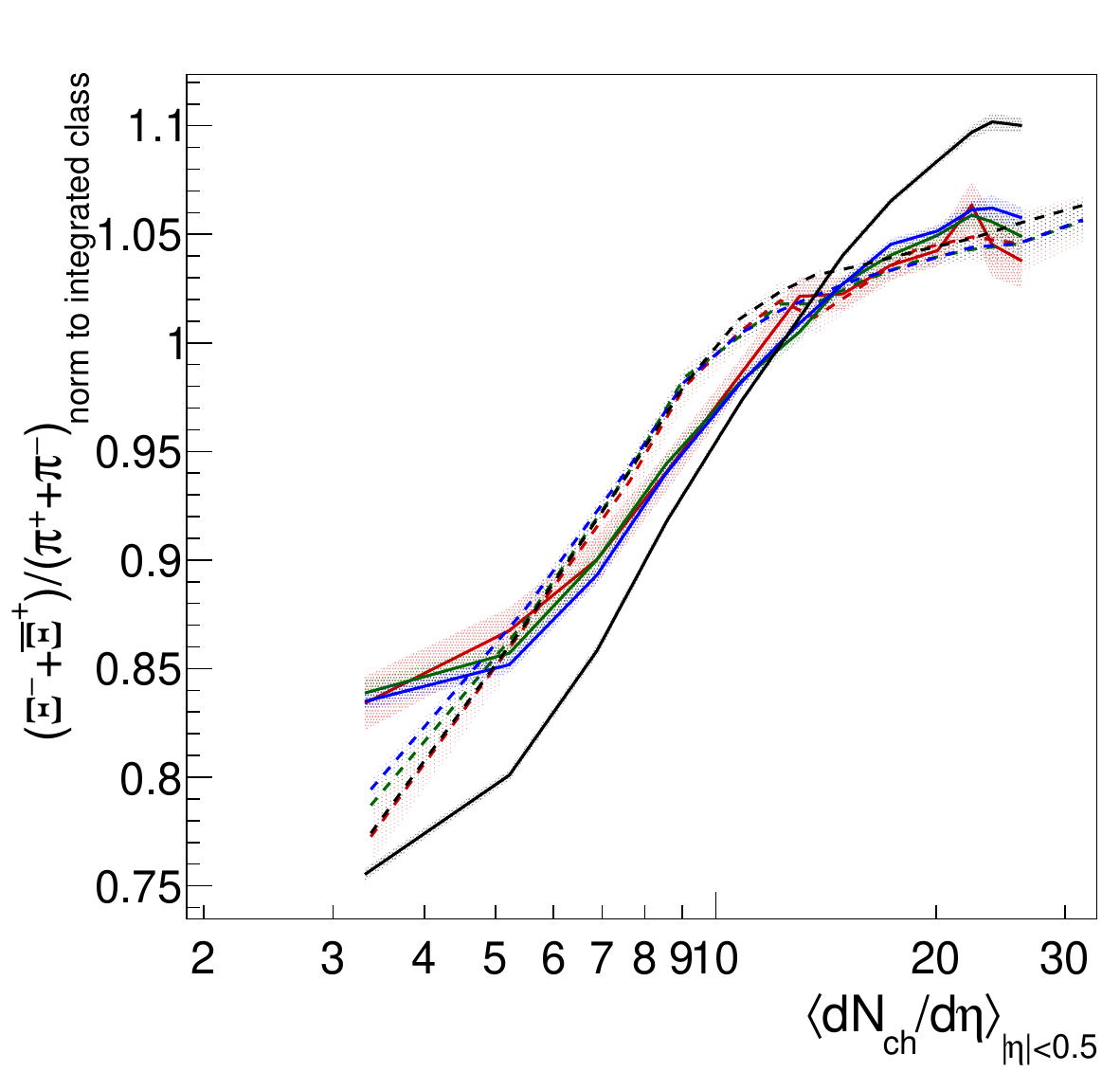}
\includegraphics[width=0.49\textwidth]{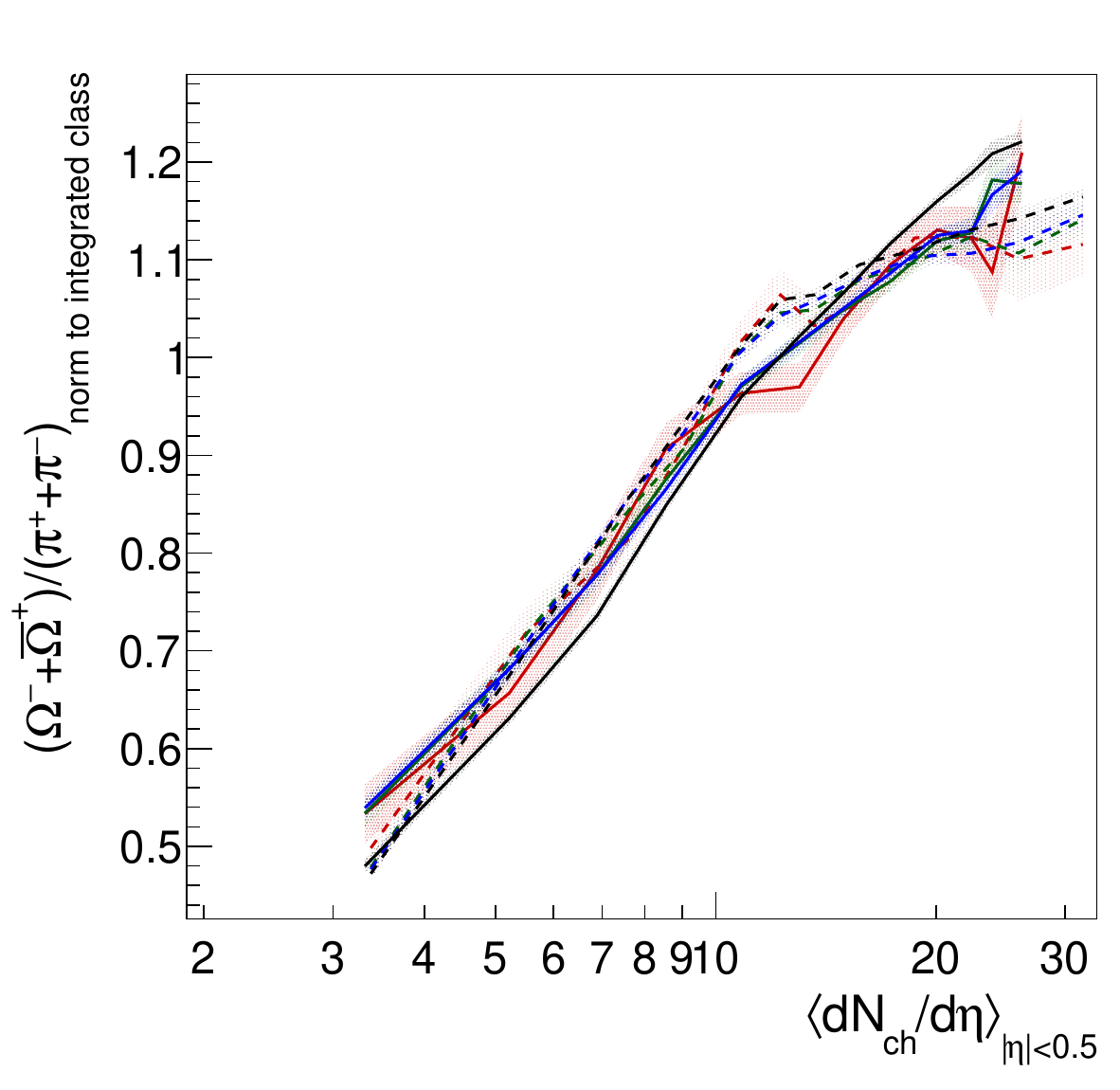}
\caption{Yield ratios of different strange hadron species measured in $|y|<2.5$ as a function of $\bigl\langle dN_{ch}/d\eta \bigr\rangle$ measured in $|\eta|<0.5$ normalised to the value integrated over all the multiplicity classes: $2K_S^0/(\pi^++\pi^-)$ (top left), $(\Lambda + \overline \Lambda)/(\pi^++\pi^-)$ (top right), $(\Xi^-+\overline \Xi^+)/(\pi^++\pi^-)$ (bottom left) and $(\Omega^-+\overline \Omega^+)/(\pi^++\pi^-)$ (bottom right).}
\label{phidoubratio}
\end{figure*}

Finally, in order to summarise the findings, and thus the suggestions for discriminating observables, we show the four hadron to pion ratios as double ratios in Fig. \ref{phidoubratio}. The double ratios are constructed as the ratio of the original curve to the value integrated over all multiplicity bins. It is clear from this visualisation, that curves for all rapidity selections overlap, indicating that the enhancement is consistent from low to high multiplicity for all the probed rapidity intervals.
Beyond that, the double ratios do not reveal many new insights, but serve as a better overview of the potential gain of performing this measurement, with all predictions brought on the same scale.

\clearpage
\appendix
\section{Event generation parameters}
\label{app:parameters}
For event generation \textsc{Pythia~8.310}\footnote{\textsc{Pythia} has been subject to minor modifications to allow for extracting the effective string tension.} and \textsc{Epos4.0.0} were employed. The choice of parameters for \pythia default are shown in Table \ref{tab:pythia-generic-parameters}, for \pythia with with ropes are shown in Table \ref{tab:pythia-ropes-parameters}. \epos parameters are shown in Table \ref{tab:epos-parameters}.

\begin{table}[h]
\centering
\caption{Parameters used to generate events with both \pythia default and with ropes.}
\label{tab:pythia-generic-parameters}
\begin{tabular*}{\columnwidth}{@{\extracolsep{\fill}}lc@{}}
\hline
 \textbf{\pythia parameters} & \\ [0.5ex]
 \hline 
 \texttt{SoftQCD:all} & \texttt{on} \\ 
 \texttt{ParticleDecays:limitTau0} & \texttt{on} \\
 \texttt{ParticleDecays:tau0Max} & \texttt{10} \\
 \texttt{333:mayDecay} & \texttt{off}\\ [0.5ex] 
 \hline
\end{tabular*}
\end{table}

\begin{table}[h]
\centering
\caption{Extra parameters used to generate events with \pythia with ropes.}
\label{tab:pythia-ropes-parameters}
\begin{tabular*}{\columnwidth}{@{\extracolsep{\fill}}lc@{}}
\hline
\textbf{\pythia with ropes parameters} & \\ [0.5ex] 
\hline
 \texttt{MultiPartonInteractions:pT0Ref} & \texttt{2.15} \\ 
 \texttt{BeamRemnants:remnantMode} & \texttt{1} \\
 \texttt{BeamRemnants:saturation} & \texttt{5} \\
 \texttt{ColourReconnection:mode} & \texttt{1} \\ 
 \texttt{ColourReconnection:allowDoubleJunRem} & \texttt{off} \\ 
 \texttt{ColourReconnection:m0} & \texttt{0.3} \\
 \texttt{ColourReconnection:allowJunctions} & \texttt{on} \\
 \texttt{ColourReconnection:junctionCorrection} & \texttt{1.2} \\
 \texttt{ColourReconnection:timeDilationMode} & \texttt{2} \\ 
 \texttt{ColourReconnection:timeDilationPar} & \texttt{0.18} \\ [0.5ex]
 \hline 
 \texttt{Ropewalk:RopeHadronization} & \texttt{on} \\
 \texttt{Ropewalk:doShoving} & \texttt{on} \\
 \texttt{Ropewalk:tInit} & \texttt{1.5} \\ 
 \texttt{Ropewalk:deltat} & \texttt{0.05} \\
 \texttt{Ropewalk:tShove} & \texttt{0.1} \\
 \texttt{Ropewalk:gAmplitude} & \texttt{0.0} \\
 \texttt{Ropewalk:doFlavour} & \texttt{on} \\ 
 \texttt{Ropewalk:r0} & \texttt{0.5} \\
 \texttt{Ropewalk:m0} & \texttt{0.2} \\
 \texttt{Ropewalk:beta} & \texttt{0.1} \\ [0.5ex]
 \hline
 \texttt{PartonVertex:setVertex} & \texttt{on} \\
 \texttt{PartonVertex:protonRadius} & \texttt{0.7} \\
 \texttt{PartonVertex:emissionWidth} & \texttt{0.1} \\ [0.5ex]
 \hline
\end{tabular*}
\end{table}

\begin{table}[h]
\caption{Parameters used to generate events with \epos.}
\label{tab:epos-parameters}
\begin{tabular*}{\columnwidth}{@{\extracolsep{\fill}}lc@{}}
\hline
 \textbf{\epos parameters} & \\ [0.5ex]
 \hline
 \texttt{MinDecayLength} & \texttt{1.0} \\ 
 \texttt{nodecays} & \texttt{331} \\
 \texttt{set ninicon} & \texttt{1} \\
 \texttt{core} & \texttt{full} \\
 \texttt{hydro} & \texttt{hlle} \\  
 \texttt{eos} & \texttt{x3ff} \\
 \texttt{hacas} & \texttt{off} \\
 \texttt{set nfreeze} & \texttt{100} \\
 \texttt{set centrality} & \texttt{0} \\ [0.5ex]
 \hline
\end{tabular*}
\end{table}

\clearpage

\bibliographystyle{spmpsci}      % mathematics and physical sciences
\bibliography{bibliography}   % name your BibTeX data base